\documentclass[twocolumn,showpacs,preprintnumbers,amsmath,amssymb,nofootinbib
,superscriptaddress]{revtex4}
\usepackage{hyperref}
\usepackage{graphicx}% Include figure files
\usepackage{color}

\newcommand{\be}{\begin{equation}}  
\newcommand{\ee}{\end{equation}}  
\newcommand{\bear}{\begin{eqnarray}}  
\newcommand{\eear}{\end{eqnarray}}  
\newcommand{\ba}{\begin{array}}  
\newcommand{\ea}{\end{array}}

\newskip\humongous \humongous=0pt plus 1000pt minus 1000pt

\newif\ifdtup

\def\oldreffmt#1{\rlap{[#1]} \hbox to 2\parindent{}}

\def\figfmt#1{\rlap{Figure {#1}} \hbox to 1in{}}  
  
\def\ie{\hbox{\it i.e.}{}}	  
\def\eg{\hbox{\it e.g.}{}}	  
\def\etal{\hbox{\it et al.}}  
  
\def\beq{\begin{equation}}  
\def\eeq{\end{equation}}  
\def\bea{\begin{eqnarray}}  
\def\eea{\end{eqnarray}}  
\def\half{\frac{1}{2}}  
  
\def\bq{\begin{quote}}  
\def\eq{\end{quote}}

\def\half{\frac{1}{2}}       
\def \etal {{\it et al.}\ }  
\relax  

\newdimen\tdim  
\tdim=\unitlength  
\def\bar{\overline}

\begin{document}
\preprint{FERMILAB-PUB-16-222-T}

{\title{Theorem:  A Static Magnetic $N$-pole Becomes
an \\
Oscillating Electric $N$-pole in a Cosmic Axion Field \footnote{Based upon invited talks given at
Oxford Universty,  ``The Axion-like Particles Workshop," IPPP, Durham University, 
U. of Minnesota, and U. of Chicago,
(spring of 2016), and ``The Invisibles Webinar," Valencia, Spain, (spring 2015).
\vspace{0.05in}} \\
}

\author{Christopher T. Hill}
\email{hill@fnal.gov}
\affiliation{Fermi National Accelerator Laboratory\\
P.O. Box 500, Batavia, Illinois 60510, USA\\$ $}

\date{\today}

\begin{abstract}
We show for the classical Maxwell equations,
including the axion electromagnetic anomaly source term,
that a cosmic axion field  induces an {\em oscillating electric $N$-moment}
for any static magnetic $N$-moment.  
This is a straightforward result, accessible to anyone who has taken a
first year graduate course in electrodynamics. 

\end{abstract}

\maketitle

%\section{Short Proof}

The action for electrodynamics in the presence of an axion field, $\theta = a(x)/f_a$, is
\cite{PDG}:
\beq
\label{action}
S =\int d^4 x\; \left(-\frac{1}{4}F_{\mu\nu}F^{\mu\nu} - \frac{1}{4} g\theta(x) F_{\mu \nu} 
\widetilde{F}^{\mu \nu} \right)
\eeq
where $\widetilde{F}^{\mu \nu}=(1/2) \epsilon^{\mu\nu\rho\sigma}F_{\rho \sigma}$.
$g$ is the anomaly coefficient, which is model dependent but typically of order $\sim 10^{-3}$.
Eq.(\ref{action}) leads to Maxwell's equations, 
$\partial_\mu F^{\mu\nu} =g\partial_\mu \theta \widetilde{F}^{\mu \nu}$.

We specialize to a cosmic axion field in its rest-frame, where it is a pure oscillator plus
a constant (or slowly varying function), $\bar{\theta}(x,t)$, 
and $m_a$ is the axion mass:
\beq
\theta(t) =\theta_0\sin(m_a (t-t_0)) +\bar{\theta}.  \qquad 
\eeq
In this frame we assume  electromagnetic fields
of the form:  $\overrightarrow{E} = \overrightarrow{E}_{r} $
and $\overrightarrow{B} = \overrightarrow{B}_{0}+\overrightarrow{B}_{r} $. 
$\overrightarrow{B}_{0}$ is a large static, applied magnetic field, and
 $\overrightarrow{E}_{r}$ and $\overrightarrow{B}_{r}$   are small oscillating 
``response fields.''\footnote{Note that $ \overrightarrow{E}_{r} $ and $ \overrightarrow{B}_{r} $ are 
of order $g\theta_0 << 1$ where $\theta_0\sim 3\times10^{-19}$ when one matches to the local
galactic halo density of $\sim 300$ MeV/cm$^3$ \cite{Hill}.}.
Maxwell's Equations in these fields, to first order in $g\theta_0 $, become: 
\bea
\label{max1}
\overrightarrow{\nabla }\times \overrightarrow{B}%
_{r}-\partial _{t}\overrightarrow{E}_{r}& = & -g\overrightarrow{%
B}_{0} \partial _{t}\theta 
\nonumber \\
 \overrightarrow{\nabla }\times \overrightarrow{E}%
_{r}+\partial _{t}\overrightarrow{B}_{r}& =& 0
\eea
\noindent
and $\ \ \overrightarrow{\nabla }\cdot \overrightarrow{B}_{r}=%
\overrightarrow{\nabla }\cdot \overrightarrow{E}_{r}=0.$
These are standard and lead to, \eg, the RF-cavity solutions \cite{Hill}

Consider an applied magnetic field, $ \overrightarrow{B}_0$, arising from
a static, classical, magnetic dipole, $\overrightarrow{m} $, with a
pointike magnetization $\overrightarrow{m}\delta^3(\overrightarrow{r})$.
This produces the familiar result,
\beq
\label{mdipole1}
\overrightarrow{B}_0= - \frac{1}{4\pi} \left( \frac{1}{r^3} \right) \left( 
\overrightarrow{m}-\frac{3\overrightarrow{r}\left( \overrightarrow{r}\cdot 
\overrightarrow{m}\right) }{r^2}\right) +\frac{2}{3}\overrightarrow{m}\delta^3(\vec{r})
\eeq
and $\overrightarrow{\nabla}\cdot\overrightarrow{B}_0 =0$.
This expression is well-known, such as in eq.(5.64)
of Jackson \cite{Jackson}. Observe, however, that eq.(\ref{mdipole1}) can be rewritten in an
equivalent form:
\beq
\label{mdipole2}
\overrightarrow{B}_0= \overrightarrow{m}\delta^3(\vec{r})+
 \frac{1}{4\pi} \overrightarrow{\nabla }\left( \overrightarrow{m}\cdot \overrightarrow{\nabla }\frac{1}{r} \right)
\eeq 
Eq.(\ref{mdipole2}) yields eq.(\ref{mdipole1}) upon computing the gradient term
by using the identity:
\beq
\label{mdipole3}
{\nabla }_i\frac{r_j}{r^3} = \delta_{ij}\frac{1}{r^3} -3 \frac{r_i r_j}{r^5}+ \frac{4\pi }{3}\delta_{ij}\delta^3(\vec{r})
\eeq
Note that, if we contract 
$(ij)$, then eq.(\ref{mdipole3}) becomes $\overrightarrow{\nabla }\cdot\frac{\vec{r}}{r^3} ={4\pi}\delta^3(\vec{r})$,
which is just Gauss' law (we emphasize that the singularities can be replaced by smooth, localized Gaussians).  The
Maxwell equations thus become: 
\bea 
\label{max2} 
\overrightarrow{\nabla }\times \overrightarrow{B}%
_{r}-\partial _{t}\overrightarrow{E}_{r}& = & -g\partial_t \theta
\left(
\overrightarrow{m}\delta^3(\vec{r})+
 \frac{1}{4\pi} \overrightarrow{\nabla }\left( \vec{m}\cdot \overrightarrow{\nabla }\frac{1}{r} \right)
 \right)
\nonumber \\
  \overrightarrow{\nabla }\times \overrightarrow{E}%
_{r}+\partial _{t}\overrightarrow{B}_{r}& = &0
\eea
We now make a redefinition of the electric field by shifting away the gradient term,
where we assume $\partial_t\bar{\theta}(x,t)\approx 0$:
\beq
\label{shift}
\overrightarrow{E}_{r}=\overrightarrow{E}'_{r}+\frac{1}{4\pi}g \widetilde{\theta} \;
\overrightarrow{\nabla }\left( \overrightarrow{m}\cdot \overrightarrow{\nabla }\frac{1}{r} \right)
\eeq
Here   $\widetilde\theta(t) =\int^t_{t_0} d\tau \partial_t \theta(t) = \theta_0\sin(m_a (t-t_0))$ 
and has the property that $\widetilde\theta(t)\rightarrow 0$ as $m_a\rightarrow 0$.
\footnote{Note that as $m_a\rightarrow 0$ then $\partial_t\theta \rightarrow 0$ and $\widetilde\theta(t)\rightarrow 0$.
The constant $\bar{\theta}$ has disappeared from the physics
and the theory has the ``shift symmetry''
$\bar{\theta} \rightarrow \bar{\theta} + k$ for any constant $k$.
This is called ``axion decoupling,'' and it is somewhat more involved with general
$\theta(x,t)$ \cite{Hill}. We note that in an RF-cavity, such as ADMX, an oscillating electric
field develops along the cavity axis driven by the axion and the same
eqs.(\ref{max1}) with $\overrightarrow{E}_{r}\propto \widetilde\theta(t)$ \cite{Hill}.
Maxwell's equations act as a ``high-pass filter'' for $\theta(t)\rightarrow\widetilde\theta(t) $.} 
The shift is a purely longitudinal (gradient) term and cannot affect the radiation field.
Since $\overrightarrow{\nabla }\times \overrightarrow{\nabla }(X) = 0$,
we have $\overrightarrow{\nabla }\times\overrightarrow{E}_{r}
= \overrightarrow{\nabla }\times\overrightarrow{E}'_{r}$,  and the second
Maxwell equation, (\ref{max2}), is unaffected by the shift.
Hence:
\bea 
\label{max3} 
\overrightarrow{\nabla }\times \overrightarrow{B}%
_{r}-\partial _{t}\overrightarrow{E}'_{r}& = & -g\partial_t \theta
\overrightarrow{m}\delta^3(\vec{r})
 \nonumber \\
  \overrightarrow{\nabla }\times \overrightarrow{E}'_{r}+\partial _{t}\overrightarrow{B}_{r}& = &0
\eea
The Maxwell equations, in terms of $\overrightarrow{B}_r$ and $\overrightarrow{E}'_{r}$,
describe the  radiation field produced by the anomaly and magnetic dipole source.
Note that the extended magnetic dipole field has completely disappeared, leaving
only the pointlike source.

Equations (\ref{max3}) are identical to those of an {\em  oscillating electric dipole moment}
$\overrightarrow{p}(t)$ (the ``Hertzian'' dipole):
 \bea 
\label{max4} 
\overrightarrow{\nabla }\times \overrightarrow{B}%
_{r}-\partial _{t}\overrightarrow{E}'_{r}& = & -\partial_t \overrightarrow{p}(t)
 \nonumber \\
  \overrightarrow{\nabla }\times \overrightarrow{E}'_{r}+\partial _{t}\overrightarrow{B}_{r}& = &0
\eea
where we have the correspondence:
\beq
\label{edipoledef}
\overrightarrow{p}(t)=g \widetilde\theta(t)
\overrightarrow{m}\delta^3(\vec{r})
\eeq
One may have noticed that, upon performing the shift, eq.(\ref{shift}), the $\overrightarrow{\nabla }\cdot \overrightarrow{E}'_{r}$  equation is now modified,  
\bea
\label{dele}
\!\!\!\!\!\!\!\! \!\!\!\!\!\!
\overrightarrow{\nabla }\cdot \overrightarrow{E}'_{r}& =& -\frac{1}{4\pi}g\widetilde\theta
\overrightarrow{\nabla }^2 \overrightarrow{m}\cdot \overrightarrow{\nabla }\frac{1}{r}
%\nonumber \\
% &=&
= g\widetilde{\theta}(t)
\vec{m}\cdot \overrightarrow{\nabla }\delta^3(\overrightarrow{r}).
\eea
In the second term,  using eq.(\ref{edipoledef}) we have:
\beq
\label{edipole1}
\overrightarrow{\nabla }\cdot \overrightarrow{E}'_{r}=\overrightarrow{\nabla }\cdot \overrightarrow{p}
\eeq
However, eq.(\ref{edipole1}) together with eq.(\ref{max4}) are precisely the Maxwell equations obtained from an action containing an electric dipole term, 
\beq
\label{edipoleaction}
S = \int d^4x \; \left(-\frac{1}{4}F_{\mu\nu}F^{\mu\nu} -\overrightarrow{p}(t)\cdot\overrightarrow{E}\right)
\eeq

The axion causes the magnetic dipole to radiate.
The radiation from a static magnetic moment in an
axion field is identical to the Hertzian electric dipole radiation (see \eg,
Jackson, \cite{Jackson} eqs.(9.18, 9.19)). In the near-zone the two terms on the {\em rhs} of eq.(\ref{shift})
cancel as $m_a\rightarrow 0$, causing $\overrightarrow{E}_{r}$ to vanish. 
Hence there is no persistent electric field in the $\widetilde{\theta}\rightarrow $ constant limit!
In the far-zone we have:
\bea
\label{rad}
\overrightarrow{E}_{r}(x,t) & = & gm_{a}^{2}\widetilde\theta(t-r)
\left( \frac{\overrightarrow{m}}{r}-\frac{%
\overrightarrow{r}}{r^{2}}\frac{\overrightarrow{m}\cdot \overrightarrow{r}}{r%
}\right) 
\nonumber \\
\overrightarrow{B}_{r}(x,t) & = & -gm_{a}^{2}\widetilde\theta(t-r)
\left( \ \overrightarrow{m}\ \times 
\frac{\overrightarrow{r}}{r^{2}}\right) 
\eea
Note the CP-violation
in the alignment of the polarization of the vector $\overrightarrow{E}_{r}$
with the axial vector $\overrightarrow{m} $, which arises from the background axion
field.  From eq.(\ref{rad}) we obtain the 
total emitted power,
\vspace{-0.20 in}
\beq
P= g^{2}m_{a}^{4}\theta_{0}^{2}|\overrightarrow{m}|^{2}/12\pi.
\eeq

\vspace{-0.3 in}
This is equivalent to the quantum result for a spin-up  to spin-up electron
with $|\overrightarrow{m}|^{2}\rightarrow \mu_{Bohr}^2 $. Coherent assemblages of many electrons
can produce potentially observable effects. 
The physical applications of this are discussed in \cite{Hill}.
In the near-zone there is a cancellation of $\overrightarrow{E}'_{r}$ with the shifted piece of eq.(\ref{shift}), so there
is no persistent constant electric dipole field in $\overrightarrow{E}_{r}$ in the $\widetilde{\theta}\rightarrow $ constant limit (see eq.(56) in \cite{Hill}).

Alternatively, we can see this result directly from the action,
 eq.(\ref{action}), without performing the shift. 
We decompose $\overrightarrow{E}_{r}$ into transverse and longitudinal (gradient) 
components:  $\overrightarrow{E}_{r}=\overrightarrow{E}_{rT} + \overrightarrow{E}_{rL}$
where (i): $\overrightarrow{\nabla}\cdot\overrightarrow{E}_{rT}=0$, and 
 (ii): $\overrightarrow{E}_{rL}= \overrightarrow{\nabla}[(1/\overrightarrow{\nabla}^2)\overrightarrow{\nabla}\cdot 
\overrightarrow{E}_{r}]$.  The anomaly term in the action of eq.(\ref{action}), 
 $\int g\theta \overrightarrow{E}_r\cdot \overrightarrow{B}_0$, 
becomes, upon integrating by parts and using (i), (ii):  $\int 
g\theta \overrightarrow{E}_{rT}\cdot \overrightarrow{m}\delta^3(\overrightarrow{r})$. Thus,   
 only the  {\em transverse electric field}
couples to the axion-induced OEDM. 
The axionic OEDM radiates transverse radiation, and it will interact in the usual way, as in eq.(\ref{edipoleaction})
with an  applied transverse $\overrightarrow{E}_T$(t),
\eg, such as a cavity mode or light. Thus, we can write a complete action 
by replacing $\overrightarrow{p}(t)$ by $\overrightarrow{p}'(t)$ in eq.(\ref{edipoleaction})
where $\overrightarrow{p}'(t)=\overrightarrow{p}(t)
-\overrightarrow{\nabla}(1/\overrightarrow{\nabla}^2)\overrightarrow{\nabla}\cdot
\overrightarrow{p}(t)$, as in  \cite{Hill2,Hill}.

We can build up a quadrupole source from a pair of  dipoles, and an
octupole source from a pair of quadrupoles, etc. Thus, by superposition, 
 the axion causes any 
static magnetic $N$-pole to become an oscillating electric $N$-pole.

We've seen that the extended magnetic dipole field is irrelevant to
the radiation, only the point-like singularity matters, a result that
translates to the $N$-pole case.  
We thus call this an {\em effective OEDM}, in the sense of \eg, Fermi's effective weak interaction.
It is ``effective''  as a leading order result in a tiny $g\theta_0$ coupling constant.

\vspace{0.1 in}
I thank W. Bardeen and A. Vainshtein for helpful discussions.
This work was done at Fermilab, operated by Fermi Research Alliance, 
LLC under Contract No. DE-AC02-07CH11359 with the United States Department of Energy.

\end{document}